\documentstyle[psfig]{mn}
	\title[Quasi-radial modes]{Quasi-radial modes of  
		rotating stars in general relativity}
	\author[Yoshida \& Eriguchi]{Shin'ichirou~Yoshida$^1$
		\thanks{yoshida@sissa.it}
		and Yoshiharu~Eriguchi$^2$\\
			$^1$SISSA, Via Beirut 2-4, 34014 Trieste, Italy\\
			$^2$Department of Earth Science and Astronomy, 
			Graduate School of Arts and Sciences, 
			University of Tokyo,\\
		 	Komaba, Meguro-ku, Tokyo 153-8902, Japan}
	\date{Accepted, Received}
\begin{document}
	\maketitle
		\begin{abstract}
		By using the Cowling approximation, quasi-radial modes 
		of rotating general relativistic stars are computed along 
		equilibrium sequences from non-rotating to maximally rotating 
		models. The eigenfrequencies of these modes are decreasing 
		functions of the rotational frequency. The eigenfrequency 
		curve of each mode as a function of the rotational
		frequency has discontinuities, which arise from 
		the {\it avoided crossing} with other curves 
		of axisymmetric modes.
	\end{abstract}


\section{Introduction}
Radial oscillations of non-rotating relativistic stars have been studied for 
over thirty years. Methods for obtaining their spectra have been 
well established (Bardeen, Thorne \& Meltzer~1966 ; see also Chapter 26 
of Misner et al.~1973 and the references therein), 
and have been applied to several equations of state (see for example 
Meltzer \& Thorne~1966). 
These works were mainly motivated by consideration of 
stellar stability 
because general relativistic effects tend to destabilize stellar 
models \cite{fwa,cs}. 

On the other hand, the effect of rotation on stellar oscillations is 
less well understood in a general relativistic context. As in the 
non-axisymmetric mode case, the slow rotation approximation has been 
the only accessible way for investigating  the eigenmode behaviour 
of rotating stars \cite{hf}. Recently, numerical relativistic hydrodynamic 
codes have
been developed by several authors and some numerical simulations of rapidly 
rotating stars have been carried out.
Stergioulas et al.~\shortcite{sfk} and Font et al.~\shortcite{fsk}
have shown that initial small perturbations around 
an equilibrium star evolved
to a superposition of normal mode oscillations
(Note that their hydrodynamic simulation is done in the fixed 
background spacetime. On the other hand, 
Shibata et al.~\shortcite{sbs} have solved the full system 
of Einstein equations to investigate the dynamical stability of
rapidly rotating stars.).

Although the excitation and evolution of these modes in realistic situations
should be investigated by time dependent hydrodynamic simulations, 
it is also important to study the mode behaviour along rotational 
equilibrium sequences by using linear perturbation theory. 

So far we have studied a few sets of non-axisymmetric eigenmodes 
of rotating stars in general relativity (f-modes by 
Yoshida \& Eriguchi~1997,1999
; p-modes by Yoshida~1999 (unpublished; presented at the 9th Yukawa
International Seminar 'BLACK HOLES AND GRAVITATIONAL WAVES 
- New Eyes in the 21st Century-') ). These results
have been obtained within the Cowling approximation 
in which Euler perturbations 
of the metric coefficients have been neglected 
(see McDermott et al.~1983 
and Finn~1988 for a definition ; 
see Lindblom and Splinter~1990 for
the accuracy of the method when applied to non-radial modes of spherical stars). 
Apart from some low order modes, these results are in good agreement 
with those of the full perturbation theory including metric perturbations
(For comparison of the eigenfrequencies for 
slowly rotating stars, see Yoshida \& Kojima~1997 
; for comparison of the neutral points of the CFS instability, 
see Yoshida \& Eriguchi~1999
which compare the results with the one obtained by 
full computation of Stergioulas \& Friedman~1998
and Morsink et al.~1999)

It is therefore natural to expect that the Cowling approximation 
could also be
successfully applied to the {\it quasi-radial} modes which are the 
smooth extensions of the radial modes of spherical stars to rotating 
stars. In the present paper, we study quasi-radial modes by using
the Cowling approximation. Contrary to the expectations, our results indicate
that {\it computations with 
this approximation can not reproduce the relativistic instability of 
spherical stars.}  This is plausible because the instability is essentially 
caused by the loss of balance between gravity and the pressure gradient, 
and in calculations of it even the small corrections of gravity 
cannot be neglected.
Moreover, the phase cancellation of the perturbed gravitational potentials, 
which may be effective in the case of non-axisymmetric modes, 
cannot be expected to happen for radial modes.

See the Appendix of the present paper for the comparison of two methods 
in the case of radial modes of non-rotating stars.

Although the validity of Cowling approximation for rotating stars is not 
fully assessed, we here {\it expect} that at least a qualitative 
picture of the eigenmode dependence 
on stellar rotation could be studied by this approximation.

\section{Results}
The equation of state used here is the polytropic one, 
$$
p = \kappa\rho^{1+\frac{1}{N}}, \quad\epsilon = \rho + Np  \eqno (1)
$$
where $\rho$, $\epsilon$ and $p$ are the rest mass density, 
energy density and pressure of 
the stellar matter, respectively. Geometrized units, 
$c=G=1$, are adopted in this paper as well as $M_\odot=1$,
following Font et al.~\shortcite{fsk}.
The constant $N$ is the polytropic index. The adiabatic exponent 
of the perturbed matter is assumed to coincide with $1+1/N$. 
The factor $\kappa$ is another constant.

Each equilibrium sequence is computed with $\kappa$ and $N$ fixed.

In the present study polar-like coordinates are used and
the metric components are written as:
$$
ds^2 = -e^{2\nu}dt^2 + e^{2\alpha}(dr^2 + r^2d\theta^2) 
+ e^{2\beta}r^2\sin^2\theta(d\phi - \omega dt)^2 .  \eqno (2)
$$
The rotational axis is located at $\sin \theta = 0$.

The coordinates used in the actual numerical computation are 
surface-fitted ones $(r^*,\theta^*)$ which are defined as:
$$
r^* = r/R_s(\theta), \quad \theta^* = \theta, \eqno (3)
$$
where $r=R_s(\theta)$ is the form of the stellar surface in equilibrium.

The numerical method used here is basically the same as that in 
Yoshida \& Eriguchi~\shortcite{ye97} where non-axisymmetric modes were 
investigated. A minor modification is needed to obtain the axisymmetric 
modes. In the case of non-axisymmetric modes, the Eulerian variable 
$\delta p/(\epsilon + p)$ is explicitly set to zero at the centre of the 
star ($\delta p$ is the Eulerian variation of the pressure). 
In the case of axisymmetric modes, however, this is not the case 
since the regularity of the solution requires 
$\partial({\delta p})/\partial{r}$ to be zero at the stellar centre. 
Therefore we simply modify the finite-difference
scheme at the innermost grid points in our numerical code. 
Moreover to avoid the coordinate singularity on the rotation axis, 
points on the axis are excluded from the computational region.  

Most of the results shown in this paper are computed 
with a resolution of 40 uniformly distributed gridpoints in the radial
$r^*$ direction and 10 in the angular $\theta^*$ one.
The computational region is a quarter of the meridional section of 
stars, thus the range of the radial and the angular coordinates are 
$0\le r^*\le 1$, $0\le\theta^*\le\pi/2$.

\begin{figure}
		\centering\leavevmode
		\psfig{file=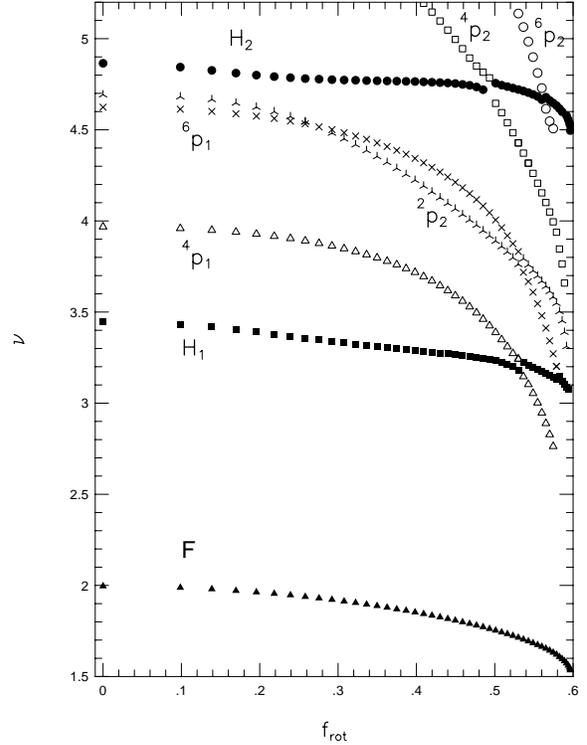,width=9.cm,angle=0,clip=}
	\caption{Sequences of axisymmetric models. Fundamental(F), 
	 first ($H_1$) and second ($H_2$) overtones 
	are plotted as well as some other axisymmetric modes.
	The polytropic parameters here are $N=0.5$, $\kappa=6.024\times 10^4$,
	$\rho_c=1.781\times 10^{-3}$. 
	Frequencies are normalized by $\sqrt{\rho_c/4\pi}$.
	The rotational frequency $f_{rot} $ at the mass-shedding limit 
	of this sequence is $0.597$ when measured in this unit. }
\label{n05freq}
\end{figure}

\begin{figure}
		\centering\leavevmode
		\psfig{file=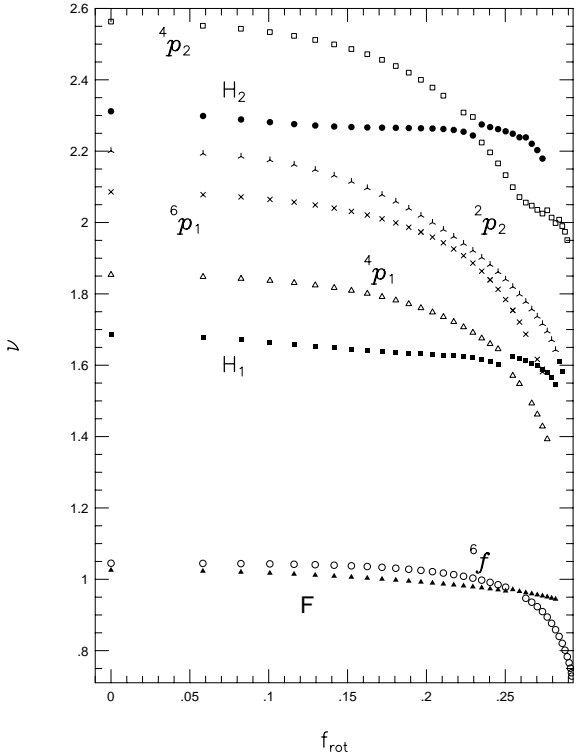,width=9.cm,angle=0,clip=}
	\caption{The same as Fig.\ref{n05freq}, except that
	the parameters for the sequence are $N=1.5$, $\kappa=4.349$, 
	$\rho_c=8.1\times 10^{-4}$.
	The rotational frequency $f_{rot} $ at the mass-shedding limit 
	of this sequence is $0.2923$ when measured in this unit. }
\label{n15freq}
\end{figure}

In Figs.~\ref{n05freq} and \ref{n15freq} 
the eigenfrequencies of the axisymmetric modes are 
plotted against the rotational frequency of the equilibrium model. 

The model parameters are tabulated in Table \ref{eqmodel}.
\begin{table}
	\caption{Parameters of the stellar model. Here $\rho_c$ 
	is the rest mass density at the stellar centre, 
	which is fixed as constant along the
	sequence. $M$ and $M/R$ are the gravitational mass and the 
	mass-to-radius ratio, where $R$ is the circumferential
	radius.}
	\begin{tabular}{cccccc}
		& $N$ & $\kappa$ & $\rho_c$ & $M/M_\odot$ & $M/R$\\
		& & & & &\\
		Fig.\ref{n05freq} & $0.5$ & $6.024\times 10^4$ & $1.781\times 10^{-3}$ & $1.4$ & $0.2$\\
		Fig.\ref{n15freq} & $1.5$ & $4.349$ & $8.1\times 10^{-4}$ & $0.57$ & $0.0564$\\
	\end{tabular}
\label{eqmodel}
\end{table}

The eigenfrequency and the rotational frequency 
are normalized by $\sqrt{\rho_c/4\pi}$, where $\rho_c$ is the
central rest mass density of the models.
The sequences '$F$', '$H_1$' and '$H_2$' are the fundamental, 
first and second overtones of the quasi-radial modes, respectively. 

Some of axisymmetric f-modes and overtones of p-modes are also plotted.
\footnote{Our numerical code assumes reflection symmetry
of the eigenmodes with respect to the equatorial plane of
the equilibrium star. As a result, modes with odd integer $l$ cannot be
computed. Considering the order of the even $l$ modes in the figure,
sequences of $^5\mbox{\it f}$ and $^3\mbox{\it p}_1$ may be located somewhere
between the the corresponding p-modes with $l=2,4$.}
These are the continuation of the corresponding f- and p-modes 
with $l=L$ and $m=0$ where $l$ and $m$ are the indices of the 
ordinary spherical harmonics $Y_{lm}(\theta,\varphi)$. 
The label $^L\mbox{\it p}_n$ refers to a mode corresponding 
to the p$_n$-mode with degree  $l=L$ 
and order $m=0$ in the non-rotating limit. 
Similarly, the mode with the label $^Lf$ is the f-mode with 
$l=L, m=0$ in the non-rotating limit.

\begin{figure}
		\centering\leavevmode
		\psfig{file=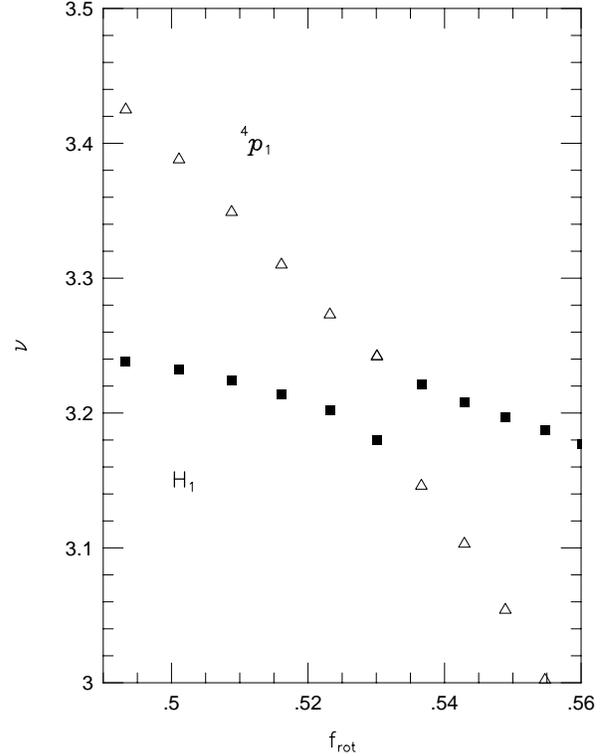,width=9cm,angle=0,clip=}
	\caption{A close-up of Fig.\ref{n05freq} around the point 
	of closest approach of the $H_1$ and $^4p_1$ sequences.}
\label{closeup}
\end{figure}

Generally the quasi-radial mode sequence encounters other sequences 
of f- or p-modes.  It is seen that the so-called {\it avoided crossing} 
occurs on these sequences (see Fig.\ref{closeup}). 
This seems to be the general 
relativistic extension of what has been found for the oscillations of 
rotating Newtonian stars. See for example Clement~\shortcite{clement} 
and Unno et al.~\shortcite{unno} for the Newtonian cases.

Fig.\ref{closeup} clearly shows the presence of the avoided crossing
whereby two eigenfrequency curves approach smoothly, 
and then depart from each 
other without crossing. At the point of closest approach, the 
characteristics of the modes on each sequence exchange. Thus the sequence 
of a mode with given characteristics has a discontinuity there. 

There are several discontinuities along a sequence whose number 
depends on the mode order as well as on the polytropic parameters of the 
equilibrium star. For the selected polytropic parameters,
discontinuities appear in the rapidly rotating models whose rotational 
frequencies are over $\sim 80~\%$ of the mass-shedding limit.

\begin{table}
 \caption{Eigenfrequencies of the lowest order quasi-radial modes.
The equilibrium sequence is constructed with the equation of state
$p=\kappa\rho^{1+1/N}$ with polytropic parameters 
$N=0.5, \kappa=6.024\times10^4,\rho_c=1.781\times 10^{-3}$. 
Here $\rho_c$ is the
baryon mass density at the stellar centre (see Table \ref{eqmodel}).
In the present case, the model is at the mass-shedding limit when 
$f_{rot}=0.597$.
The rotational frequency $f_{rot}$ and the 
eigenfrequencies are normalized by $\sqrt{\rho_c/4\pi}$. }
\begin{center}
 \begin{tabular}{cccc}
$f_{rot}$ & $F$ & $H_1$ & $H_2$\\
&&&\\
   .0000&  1.996&  3.449&  4.865\\
   .0986&  1.988&  3.433&  4.844\\
   .1696&  1.971&  3.405&  4.811\\
   .2185&  1.955&  3.379&  4.792\\
   .2581&  1.938&  3.357&  4.781\\
   .2922&  1.922&  3.339&  4.775\\
   .3224&  1.905&  3.323&  4.772\\
   .3498&  1.888&  3.310&  4.769\\
   .3749&  1.870&  3.299&  4.767\\
   .3982&  1.853&  3.288&  4.765\\
   .4301&  1.826&  3.274&  4.759\\
   .4496&  1.808&  3.265&  4.754\\
   .4679&  1.790&  3.255&  4.744\\
   .4767&  1.781&  3.250&  4.735\\
   .4851&  1.772&  3.244&  4.720\\
   .5011&  1.753&  3.232&  4.756\\
   .5088&  1.743&  3.224&  4.745\\
   .5232&  1.724&  3.202&  4.729\\
   .5301&  1.714&  3.180&  4.721\\
   .5366&  1.704&  3.221&  4.713\\
   .5429&  1.694&  3.208&  4.704\\
   .5489&  1.684&  3.197&  4.694\\
   .5601&  1.664&  3.177&  4.665\\
   .5653&  1.654&  3.166&  4.679\\
   .5701&  1.643&  3.155&  4.661\\
   .5788&  1.621&  3.130&  4.633\\
   .5826&  1.608&  3.148&  4.616\\
   .5860&  1.597&  3.120&  4.599\\
   .5916&  1.573&  3.089&  4.578\\
   .5937&  1.561&  3.076&  4.557\\

 \end{tabular}
\end{center}
\end{table}

\begin{table}
 \caption{Eigenfrequencies of the lowest order quasi-radial modes.
Polytropic parameters are $N=1.5, \kappa=4.349,\rho_c=8.1\times 10^{-4}$. 
The mass-shedding limit is $f_{rot}=0.2923$ in this case.
}
\begin{center}
 \begin{tabular}{cccc}
$f_{rot}$ & $F$ & $H_1$ & $H_2$\\
&&&\\
  0.0000&  1.027&  1.692&  2.327\\
  0.0583&  1.024&  1.684&  2.316\\
  0.0824&  1.021&  1.677&  2.306\\
  0.1007&  1.018&  1.670&  2.299\\
  0.1161&  1.015&  1.664&  2.293\\
  0.1295&  1.013&  1.659&  2.289\\
  0.1415&  1.010&  1.654&  2.287\\
  0.1524&  1.007&  1.650&  2.285\\
  0.1625&  1.004&  1.647&  2.284\\
  0.1718&  1.001&  1.644&  2.284\\
  0.1806&  0.999&  1.642&  2.283\\
  0.1965&  0.993&  1.638&  2.283\\
  0.2038&  0.990&  1.636&  2.282\\
  0.2107&  0.988&  1.634&  2.281\\
  0.2173&  0.985&  1.632&  2.279\\
  0.2235&  0.982&  1.630&  2.276\\
  0.2294&  0.980&  1.627&  2.268\\
  0.2404&  0.974&  1.619&  2.290\\
  0.2454&  0.971&  1.611&  2.284\\
  0.2502&  0.968&  1.599&  2.279\\
  0.2547&  0.964&  1.633&  2.273\\
  0.2590&  0.967&  1.626&  2.266\\
  0.2630&  0.963&  1.620&  2.256\\
  0.2668&  0.960&  1.614&  2.255\\
  0.2703&  0.957&  1.608&  2.239\\
  0.2766&  0.951&  1.591&  2.197\\
  0.2794&  0.948&  1.578&  2.248\\
  0.2841&  0.945&  1.636&  2.196

 \end{tabular}
\end{center}
\end{table}

\section{Some remarks on the analysis}

\subsection{Convergence and accuracy}
\begin{figure}
		\centering\leavevmode
		\psfig{file=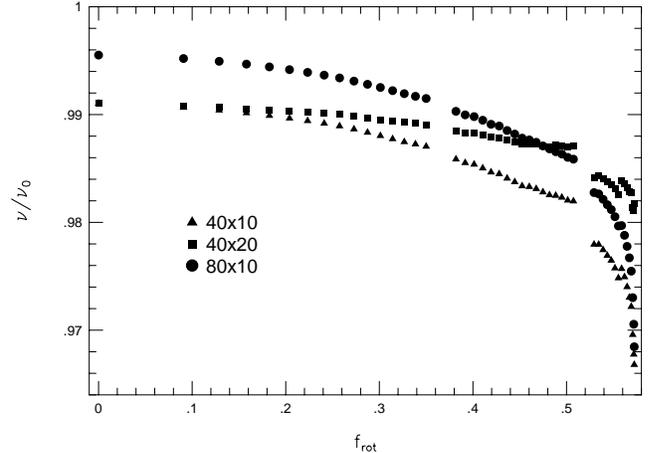,width=9.5cm,angle=90,clip=}
	\caption{Convergence of the code with increased numbers of
	grid points.
	The ratio of the eigenfrequency of the fundamental (F) mode computed 
	with given grid numbers to the extrapolated frequency 
	$\nu_0$, $\nu/\nu_0$, is plotted against 
	the rotational frequency normalized by the Kepler limit frequency.
	The polytropic parameters of the equilibrium state are $N=0.5$
	, $\kappa=10^5$ and $\rho_c=8.1\times 10^{-4}$.}
\label{convergence}
\end{figure}

Since we use a finite number of grid points, our results necessarily contain
some errors. To see how much the results
differ when the number of grid points is changed, we compute the fundamental 
quasi-radial mode sequence by using different grid numbers $(M,N) = (40,10), 
(40,20), (80,10)$ where $M$ is the number of grid points in $r^*$-direction
and $N$ is that in $\theta^*$-direction. 
By extrapolating three eigenfrequencies obtained from these grid numbers, 
$\nu_{[40,10]}, \nu_{[40,20]}, \nu_{[80,10]}$, we estimate the 
converged value of the eigenfrequency $\nu_0$ in the limit of 
infinitesimal mesh size to be:
$$
   \nu_0 = 2 (\nu_{[40,20]}+\nu_{[80,10]}) - 3 \nu_{[40,10]}.
\eqno (4)
$$
The ratio of $\nu_{[M,N]}$ to $\nu_0$ can be treated as a 
measure of convergence of the results when the grid number is changed
(Fig.\ref{convergence}). 
As can be seen from this figure, eigenfrequencies obtained by using 
$(M,N) = (40,10)$ mesh numbers (our standard resolution) agree 
with the converged values to within 3 percent.

It is noted that the relative error of $\nu_{[80,10]}$ is smaller than
that of $\nu_{[40,20]}$ in the slowly rotating cases, however for
rapidly rotating cases, the error of $\nu_{[80,10]}$ becomes larger.
This is because for the rapidly rotating cases, the deformation
of the stars from the spherical configuration 
is so large that the lack of angular resolution becomes
the main source of inaccuracy.

To check the accuracy of our two dimensional (2D) code, we have compared
the results obtained here with those obtained with 
the one dimensional (1D) code 
described in the Appendix (Table \ref{compare1d}). 
In the 1D code we employ the standard 
scheme to solve the eigenvalue problem of the linear ordinary differential 
equation within the Cowling approximation. 

As seen from this table, two results agree well to within several
percent.

\begin{table}
\caption{Eigenfrequencies of the three lowest order radial modes of 
non-rotating stars computed with the 2D Cowling code 
are compared with the results obtained with the 1D Cowling code. 
The polytropic parameters are the same as in Fig.\ref{n05freq}.
In the 2D computation, the numbers of radial and angular grid points
are $40$ and $10$, respectively.}
	\begin{tabular}{ccccc}
		& & $F$ & $H_1$ & $H_2$ \\
		Results by 1D code && $2.009$ & $3.482$ & $4.917$\\
		2D code && $1.996$ & $3.449$ & $4.865$
	\end{tabular}
\label{compare1d}
\end{table}


\subsection{Eigenfunctions and classification of modes}

Comparing with the eigenfunctions of non-axisymmetric modes, 
we notice that the eigenfunctions of quasi-radial modes 
change their shapes significantly along rotational 
equilibrium sequences. 
For example, the radial distributions of 
the Eulerian pressure perturbation and the radial component of the 
velocity perturbation change considerably near the equatorial plane of 
the star. The number of radial nodes of these functions 
increases as the stellar rotation rate increases. On the other hand, 
the overall radial dependence of the $\theta$-component of 
the velocity perturbation 
changes little as the star spins up. Therefore we can use the shape of 
this function  as a tracer of the selected mode along the equilibrium 
sequence. The nomenclature of mode sequences is based upon 
the behaviour of the mode of a non-rotating star to which the sequence is 
continued smoothly.

\begin{figure}
		\centering\leavevmode
		\psfig{file=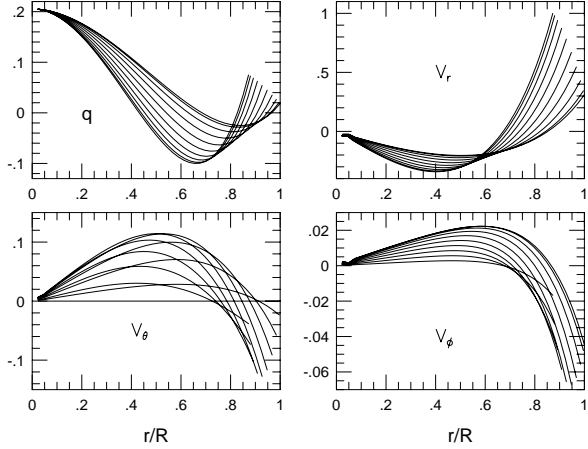,width=9.5cm,angle=90,clip=}
	\caption{Eigenfunctions of the $H_1$ mode for a slowly rotating 
	configuration.
	The polytropic parameters are the same as in Fig.\ref{n05freq}.
	The rotational frequency $f_{rot} = 0.3626$.
	Each curve shows the dependence of each eigenfunctions for a fixed
	valued of $\theta^*$
	The abscissa is the 
	radial coordinate distance normalized by the equatorial radius. 
	The curves with the largest $r$ value at their right-hand 
	end correspond to those in the equatorial plane. 
	The functions shown are $q\equiv\delta p/(p+\epsilon)$ (upper left) and 
	three velocity components $V_r$ (upper right), $V_\theta$ (lower left) 
	and $V_\phi$ (lower right), where the Eulerian perturbations
	are employed. The eigenfunctions are normalized so that 
	$V_r(r^*/R=1;\theta^*=\pi/2)=1$.
}
\label{fnH1-87}
\end{figure}

\begin{figure}
		\centering\leavevmode
		\psfig{file=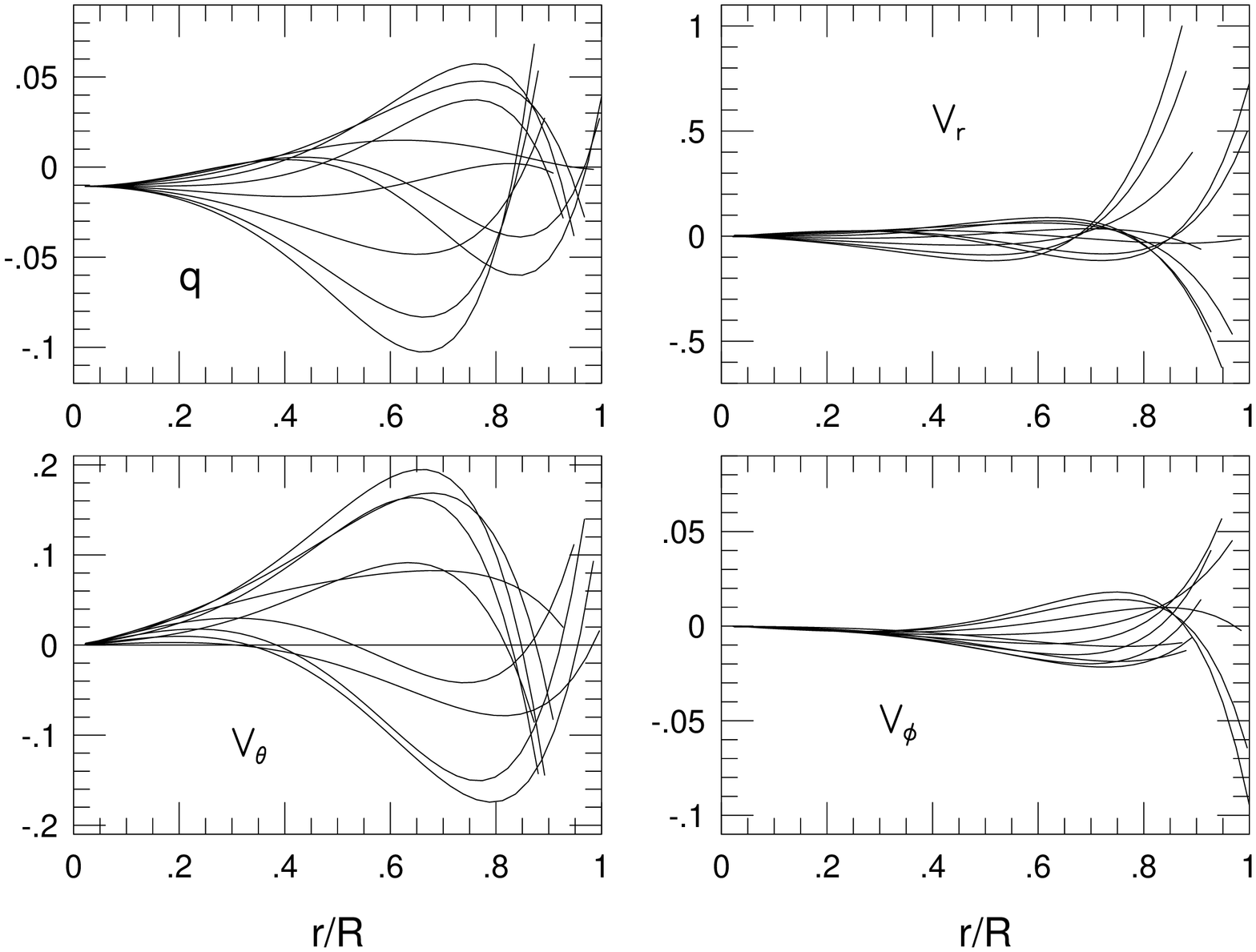,width=9.5cm,angle=90,clip=}
	\caption{Eigenfunctions of the $^4p_1$-mode for the same stellar model
	as in Fig.~\ref{fnH1-87}. 
	}
\label{fn4P1-87}
\end{figure}

\begin{figure}
		\centering\leavevmode
		\psfig{file=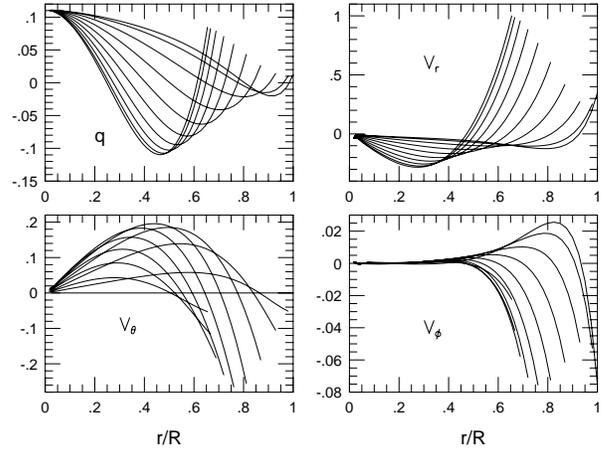,width=9.5cm,angle=90,clip=}
	\caption{Eigenfunctions of the $H_1$ mode for a rapidly rotating 
	configuration. The rotational frequency of the 
	equilibrium star is $0.5489$. }
\label{fn65-H1}
\end{figure}

\begin{figure}
		\centering\leavevmode
		\psfig{file=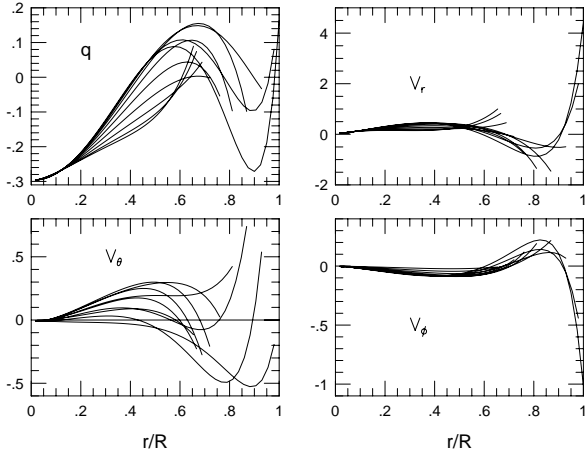,width=9.5cm,angle=90,clip=}
	\caption{Eigenfunctions of the $^4p_1$-mode for the same stellar model
	as in Fig.~\ref{fn65-H1}. }
	
\end{figure}

\subsection{Realistic neutron star models}
In addition to the polytropic case presented here, we have tried to compute 
quasi-radial modes of realistic neutron stars by using some of the candidate 
zero temperature equations of state.  
However it was rather difficult to obtain full sequences of these 
modes with our method. As the star begins to rotate, the convergence to the 
quasi-radial modes suddenly becomes much more difficult. This may partly be
due to the fact that 
these modes are sensitive (as compared with the non-radial 
modes) to the surface condition of the equilibrium star. 
Unfortunately, in the case of more realistic EOS, the adiabatic exponent
is subject to large oscillations, becoming negative in some parts. These
large variations decrease considerably the accuracy of our method which
then becomes inadequate.


\section*{Acknowledgment}
We thank Luciano Rezzolla for detailed comments on the manuscript, 
and John Miller for his help to improve it.
We also thank K\={o}ji Ury\={u} for useful discussions.


\appendix
\section{Comparison of radial modes in the full theory and in the Cowling 
approximation}

To test the validity and accuracy of the Cowling approximation
for a spherical configuration, we here present a comparison
between the results obtained by the full perturbation theory 
and by the Cowling approximation for low order radial 
modes. 
\subsection{Equilibrium model}
The space-time of the equilibrium star is characterized by the following
metric:
$$
ds^2 = -e^{2\nu(r)}dt^2 + e^{2\lambda(r)}dr^2 
+ r^2(d\theta^2 + \sin^2\theta d\varphi^2).   \eqno (A1)
$$

The metric coefficients, stellar pressure $p$ and energy density 
$\epsilon$ are obtained by integrating the standard set of equations 
with regular boundary conditions at the stellar centre:
$$
\frac{dp}{dr} = -\frac{(\epsilon + p)(m+4\pi pr^3)}{r(r-2m)}, \eqno (A2) \\
$$
$$
\frac{dm}{dr} = 4\pi r^2\epsilon, \eqno (A3)
$$
$$
\frac{d\nu}{dr} = \frac{m+4\pi pr^3}{r(r-2m)},  \eqno (A4)
$$
where $m(r)$ is defined from $e^{2\lambda} = (1-2m/r)^{-1}$.

The equation of state is assumed to be polytropic, 
$p = \kappa \rho^{1+\frac{1}{N}}$, where $\kappa$ and N are constants, 
and the rest-mass density $\rho$ is related to $\epsilon$ by 
$\epsilon = \rho + N p$.

\subsection{Equations for radial oscillations}
For radial oscillations of spherical stars, only 
the Lagrangian displacement function $\xi (r)$  
is needed to describe the physical perturbation (see Chapter 26 of 
Misner et al.~1973).

The equation of motion of the displacement ($r^2e^{-\nu}\xi\equiv\zeta$) 
in the full perturbation theory is:
$$
\zeta'' + A_f\zeta' + B_f\zeta = 0,  \eqno (A5)
$$
where $A_f$ and $B_f$ are defined as:
$$
A_f \equiv \frac{p'}{p} - \frac{2}{r} + \lambda' + 3\nu' , \eqno (A6)
$$
and 
$$
B_f \equiv \frac{\epsilon+p}{\Gamma p}\left[(\nu')^2 + 4\frac{\nu'}{r}
-8\pi p e^{2\lambda} + \sigma^2 e^{2\lambda-2\nu}\right].  \eqno (A7)
$$
The prime after a variable refers its derivative with respect to $r$.
Here $\sigma$ is the frequency and the adiabatic exponent $\Gamma$ is defined 
by:
$$
\Gamma \equiv \frac{\epsilon + p}{p}\frac{\Delta p}{\Delta\epsilon}, \eqno (A8)
$$
where $\Delta$ represents the Lagrangian perturbation. In general the adiabatic
exponent need not coincide with $1+1/N$.

In the Cowling approximation, the equation of motion is instead
($r^2e^{\lambda}\xi\equiv\eta$):
$$
\eta'' + A_c\eta' + B_c\eta = 0,  \eqno (A9)
$$
where $A_c$ and $B_c$ are defined as:
$$
A_c \equiv \frac{p'}{p} - \frac{2}{r} - \lambda' + \nu', \eqno (A10)
$$
and 
$$
B_c \equiv -\frac{p'}{\Gamma p}\left(\frac{2}{r}+\lambda'\right)
+\frac{\epsilon + p}{\Gamma p}(-\nu'' 
+ \sigma^2 e^{2\lambda-2\nu}).  \eqno (A11)
$$

These equations can be solved by the matching method: i.e.,
we have to search for $\sigma^2$ which makes the Wronskian of the
solutions, obtained by integrations from the stellar centre and from the 
surface, vanish at some matching point inside the star.

\subsection{Boundary conditions}
The boundary condition for the equation of oscillations at the centre 
of the star is regularity of the variables. It requires 
$\zeta,\eta \sim r^3$ as $r\rightarrow 0$. 

At the surface of the star, we impose the boundary and regularity 
conditions which reduce to
$$
	\zeta' + \frac{\epsilon + p}{\Gamma p'}[(\nu')^2 + 4\nu'r^{-1} 
	+ \sigma^2  e^{2\lambda-2\nu}]\zeta = 0, \eqno (A12)
$$
or
$$
	\eta' + \left[-\frac{1}{\Gamma}(2r^{-1}+\lambda') + 
	\frac{\epsilon + p}{\Gamma p'}(-\nu'' 
	+ \sigma^2 e^{2\lambda-2\nu})\right]\eta = 0. \eqno (A13)
$$

\subsection{Results}
Keeping the polytropic index $N$ fixed, 
we vary the compactness (mass-to-radius ratio) of the model 
to construct an equilibrium sequence.

In Figs.~A1 and A2  we show the typical 
sequences of the three lowest order modes. 
For all of the sequences, the eigenfrequency obtained by 
the Cowling approximation is larger than that from the full theory: i.e., 
the Cowling approximation {\it overestimates the stability} 
of the star. Before the turning point,\footnote{Note that 
with this parametrization, 
 the turning point does not correspond to the maximum mass configuration. 
 Thus the zeroes of the fundamental radial modes are not the turning points.} 
the two curves are nearly parallel. As expected, 
the relative error of the results obtained by the Cowling approximation 
becomes smaller for higher overtones. 
\begin{figure}
		\centering\leavevmode
		\psfig{file=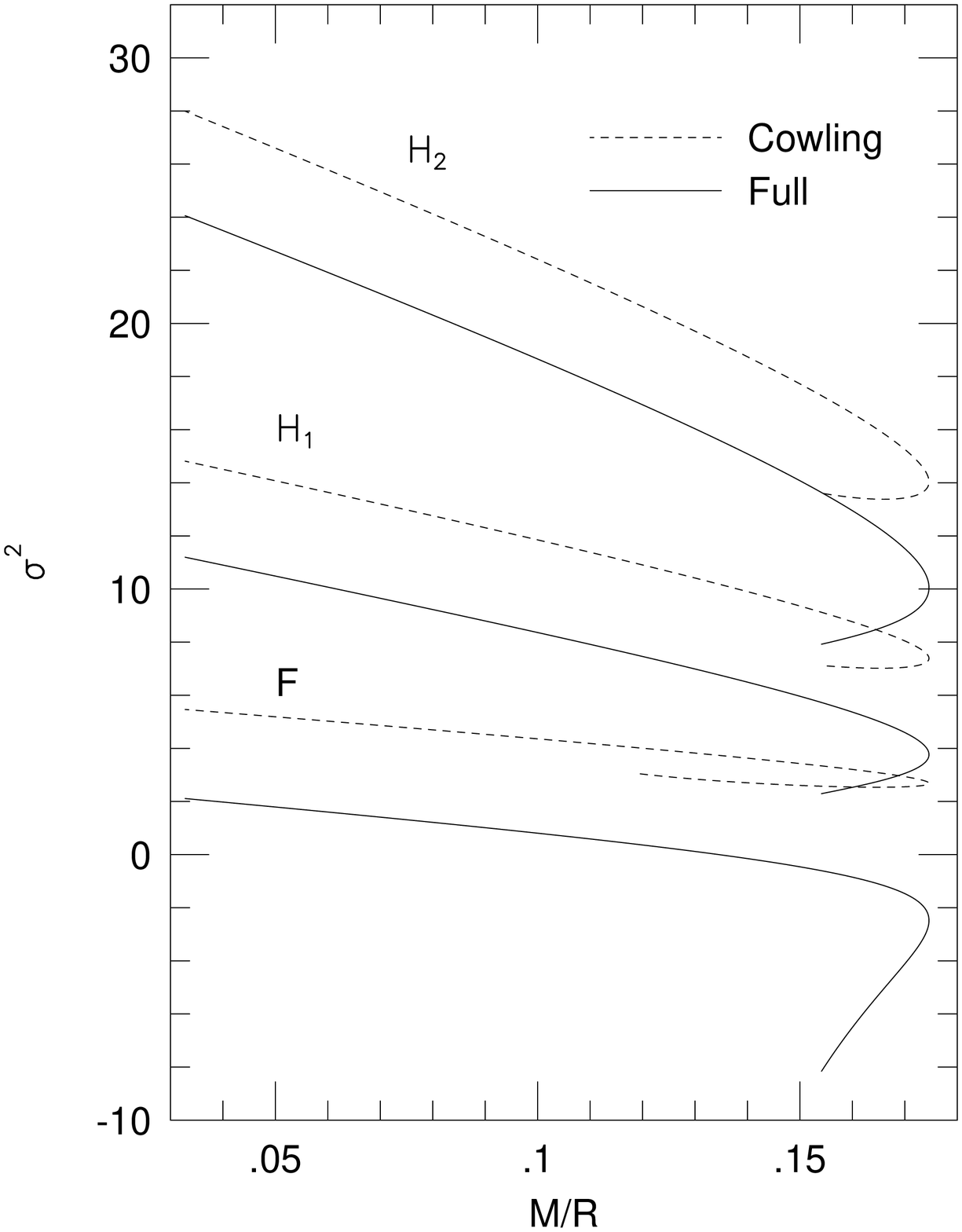,width=6.5cm,angle=0,clip=}
	\caption{Squared eigenfrequency $\sigma^2$ normalized 
	by $MR^{-3}$ is plotted as a function of the compactness $M/R$ of
	the star. Here $M$ and $R$ are the gravitational mass and the
	radius of the spherical star in the Schwarzschild coordinate, 
	respectively. Solid lines denote the eigenfrequencies of radial 
	modes in the full theory, whereas dashed lines are those for the 
	Cowling approximation. The polytropic index $N$ is $3/2$ 
	and the adiabatic exponent $\Gamma$ is $5/3$.}
\end{figure}
\begin{figure}
		\centering\leavevmode
		\psfig{file=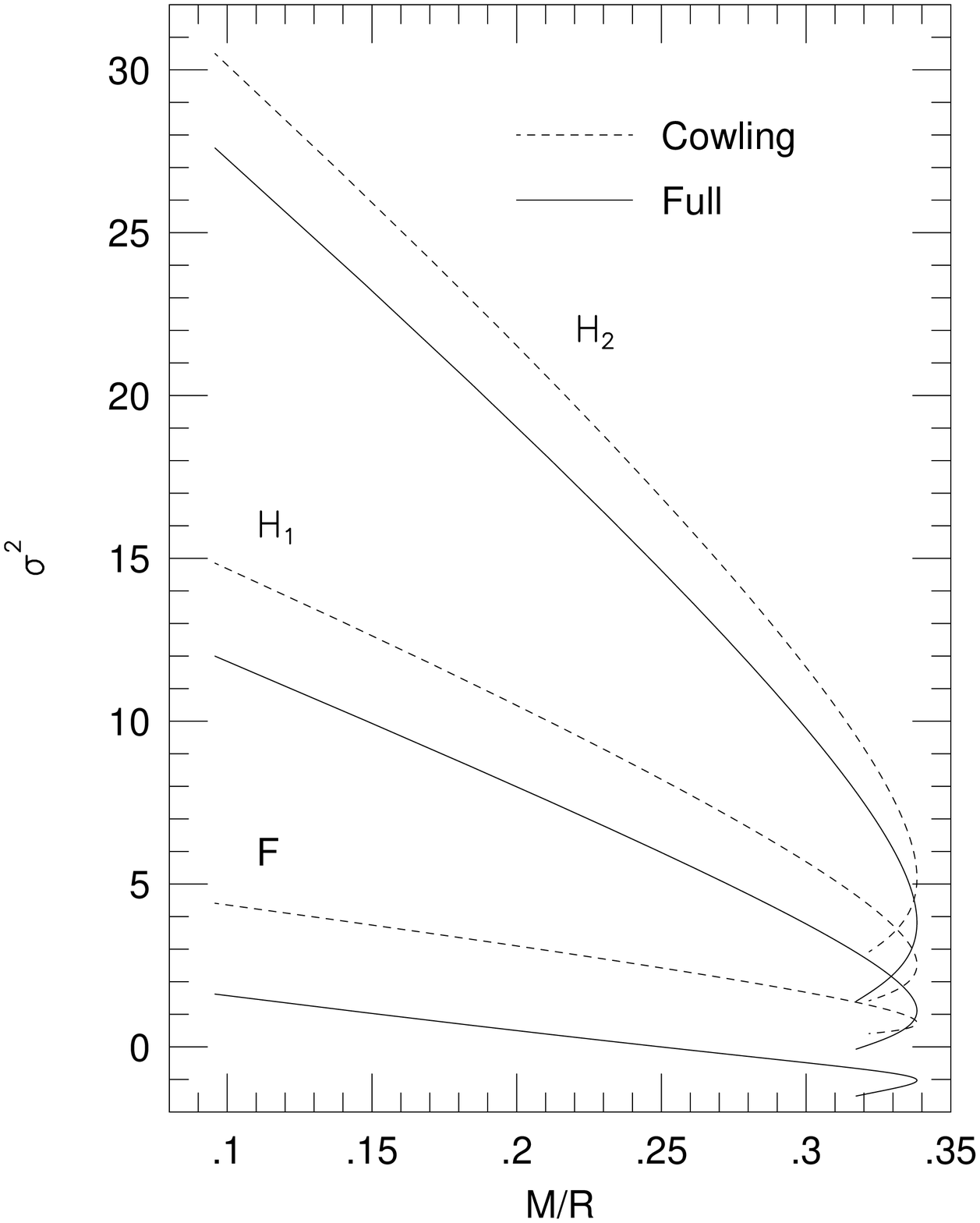,width=6.5cm,angle=0,clip=}
	\caption{The same as Fig.~A1 except that the polytropic index 
	$N=1/2$ and adiabatic exponent $\Gamma=2$. Note that $\Gamma$
	is not necessarily equal to $1+1/N$.}
\end{figure}


\begin{thebibliography}{}
	\bibitem[\protect\citename{Bardeen et al. }1966]{btm}
		Bardeen J. M., Thorne K. S., Meltzer D. W., 1966, 
		ApJ, 145, 505
	\bibitem[\protect\citename{Chandrasekhar }1964]{cs}
		Chandrasekhar S., 1964, Phys. Rev. Lett., 12, 114, 437
	\bibitem[\protect\citename{Clement }1986]{clement}
		Clement M. J., 1986, ApJ, 301, 185
	\bibitem[\protect\citename{Finn }1988]{lsf}
		Finn L. S., 1988, MNRAS, 232, 259
	\bibitem[\protect\citename{Font et al. }2000]{fsk}
		Font J. A., Stergioulas N., Kokkotas K., 2000, 
		MNRAS, 313, 678
	\bibitem[\protect\citename{Fowler }1964]{fwa}
		Fowler W. A., 1964, Rev. Mod. Phys., 36, 549 and 1104
	\bibitem[\protect\citename{Hartle \& Friedman}1975 ]{hf}
		Hartle J. B., Friedman J. L., 1975, ApJ, 196, 653
	\bibitem[\protect\citename{Lindblom \& Splinter }1990]{lis}
		Lindblom L., Splinter R. J., 1990, ApJ, 348, 198
	\bibitem[\protect\citename{McDermott et al.}1983]{mvs}
		McDermott P. N., Van Horn H. M., Sholl J. F., 1983, ApJ, 268, 837 
	\bibitem[\protect\citename{Meltzer \& Thorne }1966]{mt}
		Meltzer D. W., Thorne K. S., 1966, ApJ, 145, 514
	\bibitem[\protect\citename{Misner et al. }1973]{mtw}
		Misner C. W., Thorne K. S., Wheeler J. A., 1973, 
		Gravitation. Freeman, San Francisco
	\bibitem[\protect\citename{Morsink et al. }1999]{msb}
		Morsink S., Stergioulas N., Blattnig S. R., 1999, ApJ, 510, 854
	\bibitem[\protect\citename{Shibata et al. }2000]{sbs}
		Shibata M., Baumgarte T. W., Shapiro S. L., 2000,
		Phys. Rev., D61, 44012 
	\bibitem[\protect\citename{Stergioulas et al. }2000]{sfk}
		Stergioulas N., Font J. A., Kokkotas K., 2000, 
		in Aubourg, E., Montmerle, T., Paul, J. Peter, P., eds,
		Proceedings of the 19th Texas Symposium on Relativistic
		Astrophysics,
		Nucl. Phys. B. Proc. suppl., 80 (in CD-ROM version
		of the proceedings, numbered 07/24)
	\bibitem[\protect\citename{Stergioulas \& Friedman }1998]{sf}
		Stergioulas N., Friedman J. L., 1998, ApJ, 492, 301
	\bibitem[\protect\citename{Unno et al.}1989]{unno}
		Unno W., Osaki Y., Ando H., Saio H., Shibahashi H., 1989,
		Nonradial Oscillations of Stars, Second Edition,
		University of Tokyo Press, Tokyo
	\bibitem[\protect\citename{Yoshida \& Eriguchi }1997]{ye97}
		Yoshida S'i., Eriguchi Y., 1997, ApJ, 490, 779
	\bibitem[\protect\citename{Yoshida \& Eriguchi }1999]{ye99}
		Yoshida S'i., Eriguchi Y., 1999, ApJ, 515, 414
	\bibitem[\protect\citename{Yoshida \& Kojima }1997]{yk}
		Yoshida Sj., Kojima Y., 1997, MNRAS, 289, 117
\end{thebibliography}
\end{document}